\documentclass[10pt,conference]{IEEEtran}
\usepackage{proof}
\usepackage{amsmath}
\usepackage{amssymb}
\usepackage{stmaryrd}
\usepackage{stackengine}
\usepackage{algorithm,algpseudocode}
\usepackage{mathtools}
\usepackage{caption}
\usepackage{subcaption}

\newcommand\dldeduce{\operatorname{:-}}
\usepackage{xspace}
\newcommand{\proto}{DCV\xspace}

\newcommand{\var}[1]{\mathit{#1}}

\PassOptionsToPackage{hyphens}{url}\usepackage{hyperref}

\usepackage{xcolor}
\usepackage{listings}
\colorlet{boxcolor}{gray!10!white}
\definecolor{verylightgray}{rgb}{.97,.97,.97}

\lstdefinelanguage{Solidity}{
	keywords=[1]{anonymous, assembly, assert, balance, break, call, callcode, case, catch, class, constant, continue, constructor, contract, debugger, default, delegatecall, delete, do, else, emit, event, experimental, export, external, false, finally, for, function, gas, if, implements, import, in, indexed, instanceof, interface, internal, is, length, library, log0, log1, log2, log3, log4, memory, modifier, new, payable, pragma, private, protected, public, pure, push, require, return, returns, revert, selfdestruct, send, solidity, storage, struct, suicide, super, switch, then, this, throw, transfer, true, try, typeof, using, value, view, while, with, addmod, ecrecover, keccak256, mulmod, ripemd160, sha256, sha3}, 
	keywordstyle=[1]\color{blue},
	keywords=[2]{address, bool, byte, bytes, bytes1, bytes2, bytes3, bytes4, bytes5, bytes6, bytes7, bytes8, bytes9, bytes10, bytes11, bytes12, bytes13, bytes14, bytes15, bytes16, bytes17, bytes18, bytes19, bytes20, bytes21, bytes22, bytes23, bytes24, bytes25, bytes26, bytes27, bytes28, bytes29, bytes30, bytes31, bytes32, enum, int, int8, int16, int24, int32, int40, int48, int56, int64, int72, int80, int88, int96, int104, int112, int120, int128, int136, int144, int152, int160, int168, int176, int184, int192, int200, int208, int216, int224, int232, int240, int248, int256, mapping, string, uint, uint8, uint16, uint24, uint32, uint40, uint48, uint56, uint64, uint72, uint80, uint88, uint96, uint104, uint112, uint120, uint128, uint136, uint144, uint152, uint160, uint168, uint176, uint184, uint192, uint200, uint208, uint216, uint224, uint232, uint240, uint248, uint256, var, void, ether, finney, szabo, wei, days, hours, minutes, seconds, weeks, years},	
	keywordstyle=[2]\color{teal},
	keywords=[3]{block, blockhash, coinbase, difficulty, gaslimit, number, timestamp, msg, data, gas, sender, sig, value, now, tx, gasprice, origin},	
	keywordstyle=[3]\color{violet},
	identifierstyle=\color{black},
	sensitive=false,
	comment=[l]{//},
	morecomment=[s]{/*}{*/},
	commentstyle=\color{gray}\ttfamily,
	stringstyle=\color{red}\ttfamily,
	morestring=[b]',
	morestring=[b]"
}

\lstdefinelanguage{DSC}{
    keywords=[1]{decl,public,violation,init},keywordstyle=[1]\color{violet},
	keywords=[2]{address, bool, byte, bytes, bytes1, bytes2, bytes3, bytes4, bytes5, bytes6, bytes7, bytes8, bytes9, bytes10, bytes11, bytes12, bytes13, bytes14, bytes15, bytes16, bytes17, bytes18, bytes19, bytes20, bytes21, bytes22, bytes23, bytes24, bytes25, bytes26, bytes27, bytes28, bytes29, bytes30, bytes31, bytes32, enum, int, int8, int16, int24, int32, int40, int48, int56, int64, int72, int80, int88, int96, int104, int112, int120, int128, int136, int144, int152, int160, int168, int176, int184, int192, int200, int208, int216, int224, int232, int240, int248, int256, mapping, string, uint, uint8, uint16, uint24, uint32, uint40, uint48, uint56, uint64, uint72, uint80, uint88, uint96, uint104, uint112, uint120, uint128, uint136, uint144, uint152, uint160, uint168, uint176, uint184, uint192, uint200, uint208, uint216, uint224, uint232, uint240, uint248, uint256, var, void, ether, finney, szabo, wei, days, hours, minutes, seconds, weeks, years},	
	keywordstyle=[2]\color{teal},
    keywords=[3]{r1,r2,r3,r4,r5,r6,r7,r8,r9,r10,r11,r12,r13,r14,r2'},
    keywordstyle=[3]\color{blue},
    keywords=[4]{sum,max,count},keywordstyle=[4]\color{teal},
	comment=[l]{//},
	morecomment=[s]{/*}{*/},
	commentstyle=\color{gray}\ttfamily,
}

\lstdefinelanguage{abstract}{
    keywords=[1]{insert,search,where,on},keywordstyle=[1]\color{violet},
	keywords=[2]{address, bool, byte, bytes, bytes1, bytes2, bytes3, bytes4, bytes5, bytes6, bytes7, bytes8, bytes9, bytes10, bytes11, bytes12, bytes13, bytes14, bytes15, bytes16, bytes17, bytes18, bytes19, bytes20, bytes21, bytes22, bytes23, bytes24, bytes25, bytes26, bytes27, bytes28, bytes29, bytes30, bytes31, bytes32, enum, int, int8, int16, int24, int32, int40, int48, int56, int64, int72, int80, int88, int96, int104, int112, int120, int128, int136, int144, int152, int160, int168, int176, int184, int192, int200, int208, int216, int224, int232, int240, int248, int256, mapping, string, uint, uint8, uint16, uint24, uint32, uint40, uint48, uint56, uint64, uint72, uint80, uint88, uint96, uint104, uint112, uint120, uint128, uint136, uint144, uint152, uint160, uint168, uint176, uint184, uint192, uint200, uint208, uint216, uint224, uint232, uint240, uint248, uint256, var, void, ether, finney, szabo, wei, days, hours, minutes, seconds, weeks, years},	
	keywordstyle=[2]\color{teal},
	comment=[l]{//},
	morecomment=[s]{/*}{*/},
	commentstyle=\color{gray}\ttfamily,
}

\newcommand{\new}[1]{{\color{black}#1}}

\usepackage{stackengine}

\begin{document}

\title{Safety Verification of Declarative Smart Contracts}

\author{
\IEEEauthorblockN{
Haoxian Chen\IEEEauthorrefmark{1}, 
Lan Lu\IEEEauthorrefmark{2},
Brendan Massey\IEEEauthorrefmark{2},
Yuepeng Wang\IEEEauthorrefmark{3} 
and Boon Thau Loo\IEEEauthorrefmark{2}} 
\IEEEauthorblockA{
\IEEEauthorrefmark{1}ShanghaiTech University, 
\IEEEauthorrefmark{2}University of Pennsylvania,
\IEEEauthorrefmark{3}Simon Fraser University\\
hxchen@shanghaitech.edu.cn,
\{lanlu,masseybr,boonloo\}@seas.upenn.edu,
yuepeng@sfu.ca
}
}

\maketitle

\begin{abstract}

Smart contracts manage a large number of digital assets nowadays.
Bugs in these contracts have led to significant financial loss.
Verifying the correctness of smart contracts is, therefore, an
important task.
This paper presents an automated safety verification tool, \proto,
that targets declarative smart contracts written in DeCon,
a logic-based domain-specific language for smart contract
implementation and specification.
\proto proves safety properties by mathematical induction 
and can automatically infer inductive
invariants using heuristic patterns, without annotations from the developer.
Our evaluation on 20 benchmark contracts
shows that \proto is effective in verifying smart contracts
adapted from public repositories, and can verify contracts not supported by other tools. Furthermore, \proto significantly outperforms
baseline tools in verification time.

\end{abstract}
\section{Introduction}

Smart contracts are programs that process transactions on blockchains -- a
type of decentralized and distributed ledgers.
The combination of smart contracts and blockchains has enabled a wide range of
innovations in many fields including banking~\cite{erc20-tokens}, 
trading~\cite{notheisen2017trading,nft}, 
and financing~\cite{wang2018overview}, etc.

Nowadays, smart contracts are managing a massive amount of digital assets 
\footnote{According to \href{https://etherscan.io/tokens}{Etherscan}, as of the writing of this paper,
the top ERC20 tokens are managing billions of dollars worth of tokens.},
but they also suffer from security vulnerabilities~\cite{dao,dice2win,kingofether},
which have lead to significant financial loss.
In addition, since smart contracts are stored and executed on 
blockchains, once they are deployed, it is hard to terminate execution 
and update the contracts when new vulnerabilities are discovered.

One way to reduce potential vulnerabilities, whose patterns are unknown
during development, is safety verification:
a smart contract that is always safe is less likely to suffer from 
undiscovered vulnerabilities\cite{kalra2018zeus,permenev2020verx}.
Thus, this work focuses on the problem of safety verification for smart contracts.

Most existing solutions directly verify the implementation of smart 
contracts~\cite{solidity-smt,hajdu2019solc,marescotti2020accurate,kalra2018zeus,permenev2020verx}.
These solutions have worked very well on verifying 
properties concerning one single transaction,
e.g., 
integer overflow. However, when it comes to 
properties that needs to hold across an infinite sequence of transactions,
these approaches suffer from low efficiency due to state explosion issues.
Some solutions~\cite{frank2020ethbmc,antonino2020formalising} 
trade soundness for efficiency, verifying properties up to a certain number of transactions.

On the other hand, in model-based verification approaches,
a formal model of the smart contract is specified
separately from the implementation.
Given such a formal model and the implementation, two kinds of verification problems are studied: 
(1) Does the formal model satisfy the desired properties~\cite{nehai2018model,cassez2022deductive}?
(2) Is the implementation consistent with the formal model~\cite{chen2018language}?
While this verification approach is generally more efficient, 
as the formal model abstracts away implementation details that are 
irrelevant to the verification task, 
it does require additional effort from the user 
since they need to specify the formal model in addition to the implementation. 
Additionally, the steep learning curves of formal specification 
languages may limit the adoption of such verification approach.

This paper proposes an alternative verification approach based on
an executable specification of smart contracts. 
In particular, we target smart contracts written in DeCon~\cite{chen2022declarative},
a domain-specific language for smart contract specification and implementation.
A DeCon contract is a declarative specification for the smart contract by itself,
making it more efficient to reason about than the low-level implementation in Solidity.
It is also executable, in that it can be automatically compiled
into a Solidity program which can be deployed and run on the Ethereum blockchain.
Automatic code generation based on the verified specification 
can save developers the manual effort of implementing the contract.
The high-level abstraction and executability of DeCon make it 
an ideal target for verifying contract-level properties.

We implement a prototype, \proto (\underline{D}e\underline{C}on \underline{V}erifier), 
for verifying declarative smart contracts. 
Properties are specified as declarative queries for safety violations in the DeCon
language.
\proto verifies safety invariants using mathematical induction
on the sequence of transactions.
A typical challenge in induction is to infer inductive invariants that can help
prove the target property.
Our key insight for addressing this challenge is that the DeCon language exposes the exact logical
predicates that are necessary for constructing inductive invariants,
which makes inductive invariant inference tractable.

Another benefit of using DeCon is that 
it provides uniform interfaces for both contract implementation 
and property specification.
Specifically, DeCon models the smart contract states as relational databases, and properties as
violation queries against these databases. 
Thus, developers can specify both the contract logic and its properties 
in a declarative and succinct way,
and finish the verification and implementation automatically.

This paper makes the following contributions.
\begin{itemize}
    \item A verification method for smart contracts,
      targeting contract-level safety invariants based on a declarative specification language and the induction proof strategy 
      (Sections~\ref{sec:program-transformation},~\ref{sec:verification-method}).
    \item A domain-specific adaptation of the Houdini algorithm~\cite{lahiri2009complexity} 
        to infer inductive invariants for automated proof (Section~\ref{sec:verification-method}).
    \item An open-source verification tool for future study and comparison.
    \item Evaluation that compares \proto with state-of-the-art verification
      tools, on 20 representative benchmark smart contracts. 
      Specifically, \proto successfully verifies all benchmarks, including the ones not supported by other tools. Furthermore, it is significantly more efficient than other tools in terms of verification time 
      (Section~\ref{sec:eval}).
\end{itemize}

\section{Illustrative Example}

\begin{figure}[ht]
    \centering
    \includegraphics[width=0.48\textwidth]{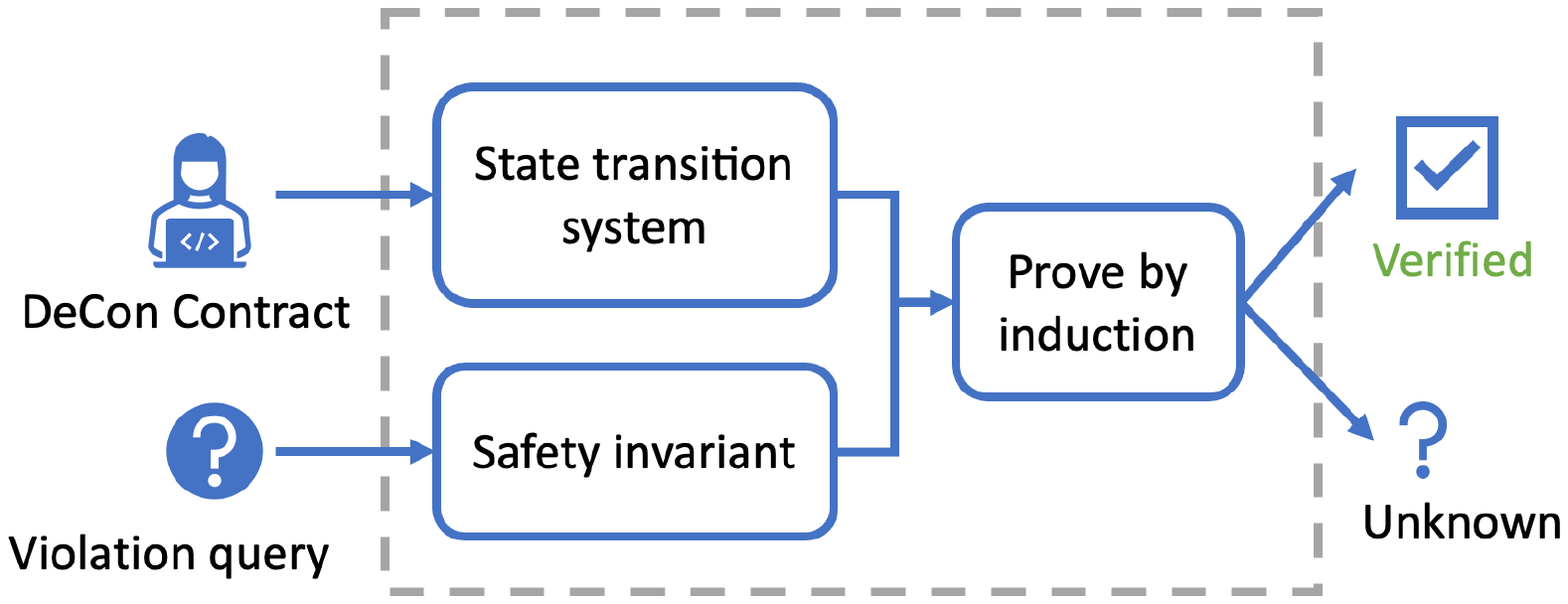}
    \caption{Overview of \proto.}
    \label{fig:overview}
\end{figure}

Figure~\ref{fig:overview} presents an overview of \proto. 
It takes a smart contract and a property specification
(in the form of a violation query) as input,
both of which are written in the DeCon language (Section~\ref{sec:language}).
The smart contract is then translated into a state transition
system, and the property is translated into a safety invariant
on the system states.
\proto then verifies the transition system preserves the safety 
invariant by mathematical induction. 
In our prototype, the verification is performed by Z3~\cite{z3},
an automated theorem prover.

If the verification succeeds, 
\proto guarantees that the smart contract is safe 
by ensuring that the violation query result is always empty, 
and returns an inductive invariant as a proof. 
However, if the verification fails, 
\proto returns ``unknown'', 
indicating that the smart contract may not satisfy the specified safety invariant.

In the rest of this section, we use a voting contract 
as an example to illustrate the workflow of \proto. 
This example is adapted from the voting example in Solidity~\cite{solidity-examples}, 
simplified for ease of exposition.

\subsection{A Voting Contract}
\label{sec:decon-example}

\begin{figure}[ht]
    \centering
    \includegraphics[width=0.5\textwidth]{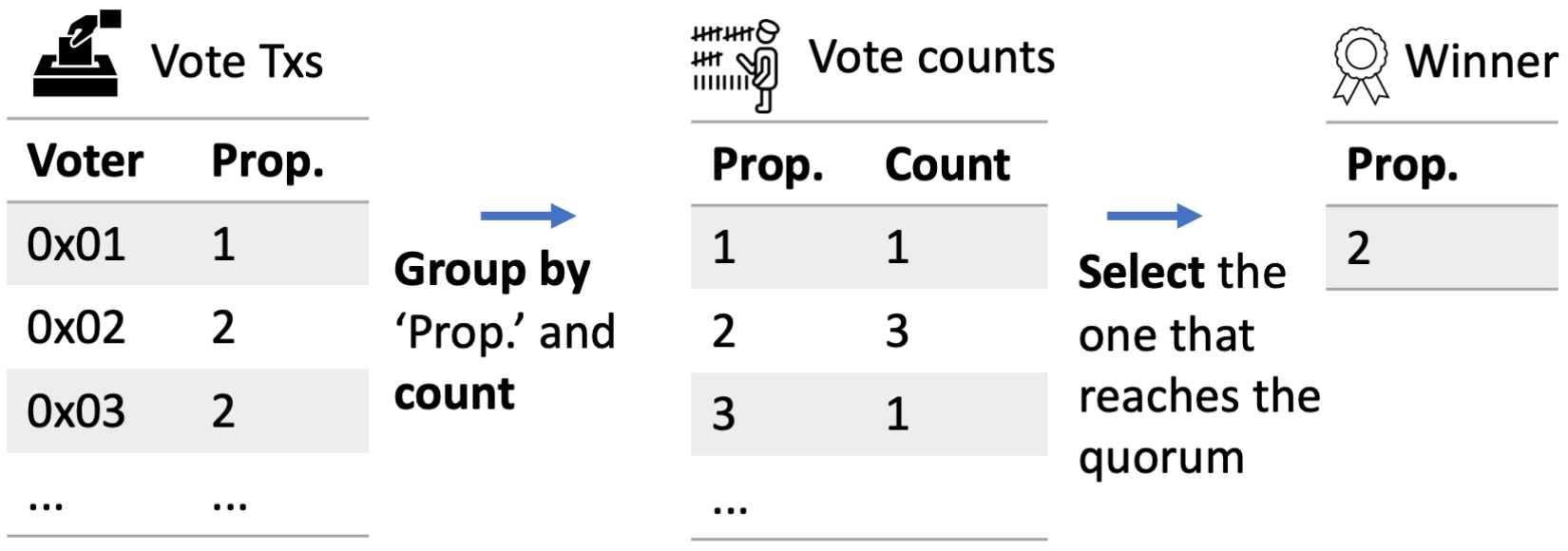}
    \caption{A voting contract}
    \label{fig:example}
\end{figure}

\begin{figure}[ht]
\centering
\begin{lstlisting}[
caption={A smart contract for voting, written in
    DeCon~\cite{chen2022declarative}.},
label={lst:voting}, 
frame=single,
numbers=left,
xleftmargin=1.5em,
% framexleftmargin=1.5em,
numbersep=5pt,
language=DSC,
]
/* Declare relations. */
.decl recv_vote(proposal: uint)
.decl vote(p: address, proposal: uint)
.decl isVoter(v: address, b: bool)[0]
.decl votes(proposal: uint, c: uint)[0]
.decl wins(proposal: uint, b: bool)[0]
.decl voted(p: address, b: bool)[0]
.decl *winningProposal(proposal: uint)
.decl *hasWinner(b: bool)
.decl *quorumSize(q: uint)

.init isVoter

/* Transaction where voter v cast a vote to proposal p. */
vote(v,p) :- recv_vote(p), msgSender(v), 
             hasWinner(false), voted(v, false), 
             isVoter(v, true).

/* Count votes for each proposal p. */
votes(p,c) :- vote(_,p), c = count: vote(_,p).

/* A proposal wins by reaching a quorum. */
wins(p, true) :- votes(p,c), quorumSize(q), 
                 c >= q.
hasWinner(true) :- wins(_,b), b==true.
winningProposal(p) :- wins(p,b), b==true.

voted(v,true) :- vote(v,_).

/*Safety invariant: at most one winning proposal.
*/
.decl inconsistency(p1: uint, p2: uint)[0,1]
.violation inconsistency
inconsistency(p1,p2) :- wins(p1,true),
                        wins(p2,true),p1!=p2.
\end{lstlisting}
\end{figure}

\begin{figure*}
    \centering
    \includegraphics[width=0.8\textwidth]{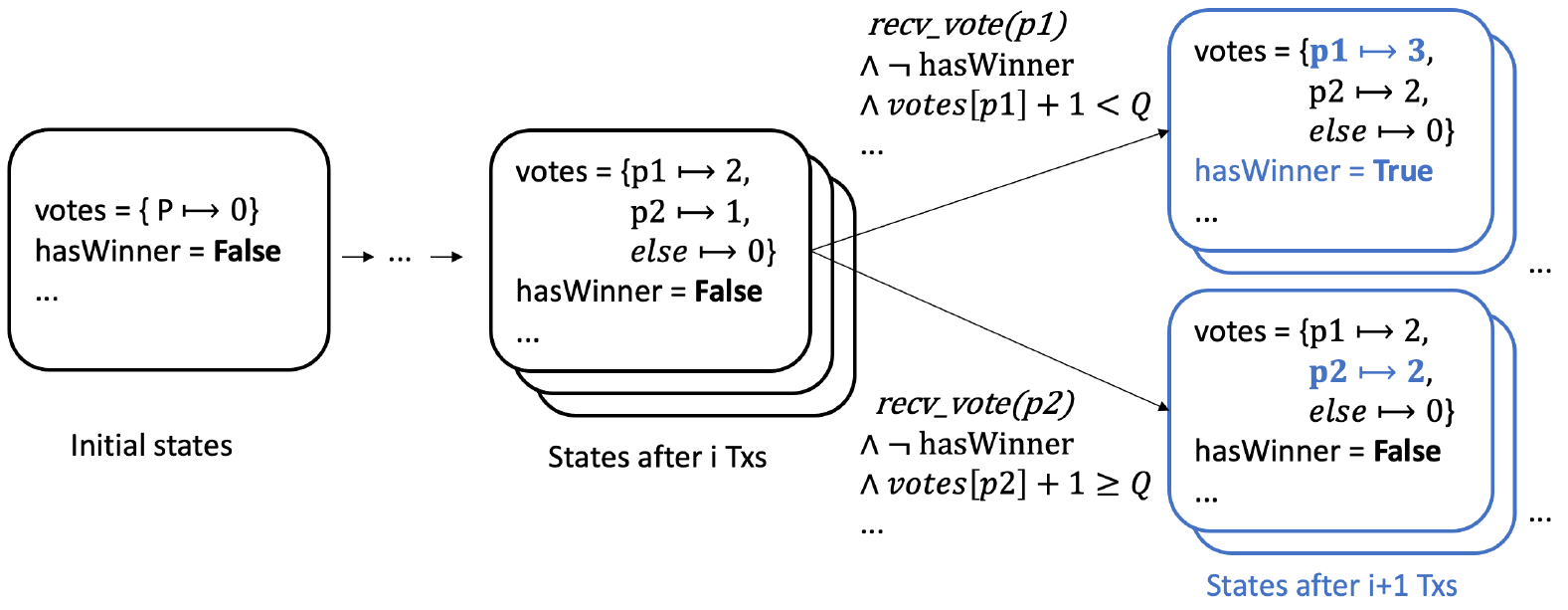}
    \caption{The voting contract as a state transition system.}
    \label{fig:transition-system}
\end{figure*}

Figure~\ref{fig:example} illustrates a voting scenario in a declarative 
view (i.e., everything is represented as a relational table):
\begin{enumerate}
    \item 
    Participants cast their votes by sending \lstinline{vote} transactions, 
    and these transaction records are stored in the ``Vote Txs'' table (on the left), 
    with the voter address and proposal ID (``Prop.'') listed as columns.
    \item 
    For each proposal $p$, the number of votes it receives can be counted 
    by grouping the entries in the ``Vote Txs'' table by the ``Prop.'' column,
    and then counting the number of entries within each group. 
    The counting results are displayed in the ``Vote counts'' table (middle).
    \item 
    The proposal that first reaches a quorum is declared as the winner. 
    Suppose in this example there are 5 participants and the quorum size is 3,
    proposal 2 is selected as the winner as it gets 3 votes.
\end{enumerate}

\subsection{Smart contract written in DeCon language}

Listing~\ref{lst:voting} shows the implementation of this voting
contract in DeCon~\cite{chen2022declarative},
which consists of three major components:

\noindent \textbf{(1) Relation declaration and annotation.}
The relations shown in Figure~\ref{fig:example}, along with other auxiliary relations, 
are declared in lines 1 to 10 of Listing~\ref{lst:voting}. 
These declarations define the table schema in relational databases, 
where each schema consists of the table name followed by column names and types in parentheses. 
Optionally, a square bracket annotates the index of the primary key columns, 
indicating that these columns uniquely identify a row. 
For example, the relation \lstinline{votes(proposal: uint, c: uint)[0]} 
on line 5 has the first column, proposal ID, as the primary key 
because votes are counted for each unique proposal. 
If no primary keys are annotated, 
all columns are interpreted as primary keys, 
meaning that the table is a set of tuples.

A special kind of relation is a singleton relation, annotated by $*$. 
Singleton relations only have one row, 
such as \lstinline{winningProposal} in line 8.

By default all relational tables are initialized to be empty, except relations
annotated by the \lstinline{init} keyword (line 12).
These relations are initialized by the constructor arguments passed during deployment.

\noindent \textbf{(2) Relation definition in inference rules.}
Each relation is defined in the form of a rule, \lstinline{head :- body.}
Similar to the rules used in Datalog programs,
the body consists of a list of relational literals, and is evaluated
to true if and only if there exists a valuation of all variables 
such that each literal has the corresponding concrete entries
in the table.
If the body is true, the head is inserted into the corresponding table.

For instance, the rule in line 15 specifies that a \lstinline{vote} transaction
can be committed if there is no winner yet, 
the message sender is a registered voter, and the voter has not voted before.
The literal \lstinline{recv_vote(p)} represents a transaction handler
that evaluates to true upon receiving a \lstinline{vote} transaction request.
Rules that contain such transaction handlers (literal with a \lstinline{recv_}
prefix in the relation name) are referred to as transaction rules.
Committing a transaction inserts a new entry into the transaction table 
(``Vote Txs'' in Figure~\ref{fig:example}).

Inserting a new \lstinline{vote(v,p)} literal also triggers updates to all its 
direct dependent rules.
A rule is considered directly dependent on a relation $R$ if and only
if a literal of relation $R$ is in its body.
In this case, relation \lstinline{votes} and \lstinline{voted} are updated next.
The chain of dependent rule updates continues until no further dependent rules 
can be triggered, and the transaction handling is finished.
Using this mechanism, the votes for each proposal, as well as the winning proposal, 
are automatically updated when new votes are approved.

On the other hand, 
if the body of a transaction rule evaluates to false upon receiving a transaction request, 
the transaction request is rejected, 
and no updates are made to any of the affected relations.

\noindent \textbf{(3) Properties as violation query.}
Line 32 specifies a safety property as another relation,
which is further annotated as a violation query in line 33.
This relation is defined by the rule in line 34.
If the rule is evaluated to true, it means that there exists
two different winning proposals, indicating a violation to the safety invariant
that there is at most one winning proposal.
Such violation query rule is expected to be always false during the execution
of a correct smart contract.

\subsection{Translating DeCon Contract to State Transition System}

In order to perform formal verification,
DeCon contracts are encoded as state transition systems.
The state space comprises all possible valuations
of the relational tables, and each successful transaction
triggers an atomic state transition step
(the transaction atomicity is guaranteed by the underlying 
Ethereum blockchain).
Such encoding naturally captures the semantics of smart contracts: 
reactive programs that listen and respond to requests (transactions).

Figure~\ref{fig:transition-system} illustrates part of the transition system translated
from the voting contract in Listing~\ref{lst:voting}.
The middle portion (labeled ``States after i Txs'')
shows a state that is reached after $i$ transactions from one of the initial states.
At this point in the execution, proposal $p_1$ has received two votes, 
proposal $p_2$ has one, and no winner has been declared yet.
Two outgoing edges from this state are highlighted.
The one on top represents a \lstinline{vote(p1)} transaction,
where $p_1$ receives an additional vote, 
thereby achieving the quorum and becoming the winning proposal. 
This transaction can be executed only if certain conditions are met, 
which are annotated on the edge
(only part of the conditions are shown due to space limit).
The edge is derived from the transaction rule $r$ in Listing~\ref{lst:voting} line 15
($\var{recv\_vote}(p_1) \land \neg hasWinner \land ...$),
and its dependent rules from line 19 to 26 ($\var{votes}[p_1] \geq Q \land ...$).
This edge leads to a new state in which the number of votes for proposal $p_1$ is incremented by one, 
and it becomes the winner, which is also translated from line 19 to 26.

Similarly, the bottom right shows another transaction where proposal $p_2$ gets
a vote, but $\var{hasWinner}$ remains $False$ since no proposal has reached the quorum.

Section~\ref{sec:transition} formally describes the algorithm to 
translate a DeCon smart contract into a state transition system.

\noindent \textbf{Property.}
The violation query rule (line 31) is translated into the following safety invariant:
\begin{equation}
\label{eq:consistency}
\neg [ \exists p_1,p_2.~ wins(p1) \land wins(p2) \land p1 \neq p2 ]
\end{equation}
It states that there do not exist proposals $p1$ and $p2$ such
that the violation query is true,
which means there is at most one winning proposal.

\subsection{Proof by Induction}

To prove safety invariants of a smart contract against an infinite sequence 
of transactions, \proto adopts the mathematical induction approach.
Given a state transition system, and a safety invariant,
the proof consists of two steps:
\begin{enumerate}
    \item Base case: all initial states satisfy 
    the safety invariant.
    \item Induction step: if the safety invariant holds
    for some state $s$, then it also holds for all possible next states $s'$.
\end{enumerate}

One of the biggest challenges in automatic induction proof
is finding inductive invariants.
In some cases, a true safety invariant may not be inductive,
which means that although the safety invariant is true for all possible states,
it can still fail the induction proof step.
To successfully complete the induction step,
an inductive invariant $inv(s)$ needs to be found, 
such that $inv(s) \land prop(s) $ is inductive, 
where $s$ is the state variable.

For example, the safety invariant in Equation~\ref{eq:consistency}
cannot be proved inductively on its own.
To make it inductive,
it needs to be augmented by an inductive invariant, such as the following:
\begin{equation}
    \label{eq:invariant-example}
    \forall u \in \var{Proposal}. ~ \var{wins}[u] \implies \var{hasWinner}
\end{equation}
which asserts that if any proposal $u \in \var{Proposal}$ is marked 
as the winner, the predicate $hasWinner$ must also be true.

Inductive invariants are typically inferred in
a guess-and-check manner~\cite{lahiri2009complexity},
where a set of candidate invariants are enumerated until
an inductive one is found.
However, such approaches heavily rely on good heuristics to generate a 
set of candidate invariants.
The insight of \proto is that, 
rules in DeCon contracts provide a concise and high-quality source of logical predicates 
for constructing inductive invariants.
Since they only concern high-level logic 
and do not include implementation details, 
the extracted predicates are much smaller in size than 
those extracted from lower-level implementations, 
greatly speeding up the invariant search process.
For instance, the predicates in Equation~\ref{eq:invariant-example}
($\var{wins}[u]$, $\var{hasWinner}$),
are presented in the DeCon transaction rules in Listing~\ref{lst:voting}.
We describe the details of predicate extraction and inductive invariant generation
in Section~\ref{sec:verification-method}.

\section{The DeCon Language}
\label{sec:language}

\begin{figure}[ht]
\[
\boxed{
\begin{array}{rcl}
    Decl & \coloneqq & .decl ~ R\\
    Annot & \coloneqq & .(init ~|~ violation ~|~ public)~R  \\ 
    (Relation)~R & \coloneqq & SR ~|~ SG ~|~ TR \\
    (Simple)~SR & \coloneqq & .decl~Str(Schema)[K] \\
    (Singleton)~SG & \coloneqq & .decl~*Str(Schema) \\
    (Transaction)~TR & \coloneqq & .decl~*recv\_Str(Schema) \\
    (Primary~keys)~K & \coloneqq & [k1,k2,...] \\
    Schema & \coloneqq & Str: T1,~Str: T2, ... \\
    (Type)~T & \coloneqq & address ~|~ uint ~|~ int ~|~ bool \\
\end{array}
}
\]
\caption{Syntax of DeCon relation declaration and annotation.}
\label{fig:syntax}
\end{figure}

A DeCon contract consists of three main blocks:
(1) Relation declarations, (2) Relation annotations, and (3) Rules.
\[
\boxed{
    (Contract)~P \coloneqq Decl ~|~ Annot ~|~ Rule 
    }
\]

\noindent \textbf{Relation declarations.}
As shown in Figure~\ref{fig:syntax},
there are three kinds of relation declaration syntax:
\begin{itemize}
    \item
    Simple relations ($SR$) have a string for a relation name,
    followed by a schema in parenthesis, and optional primary 
    key indices in a square bracket.
    The schema consists of a list of column names and types.
    When inserting a new tuple to a table, 
    if a row with the same primary keys exists, then the row is replaced by the new tuple.

    \item 
    Singleton relations ($SG$) are relations annotated with a \lstinline{*} symbol.
    These relations have only one row.
    Row insertion is also an update for singleton relations.

    \new{
    \item 
    Transaction relations ($TR$) are relations with prefix $recv\_$,
    which are interpreted as an event trigger for incoming
    transaction requests.
    For example, in line 15 of listing~\ref{lst:voting}, 
    the literal \lstinline{recv_vote(p)} is triggered to be true
    when the contract receives a \lstinline{vote} transaction,
    with parameter \lstinline{p} being the proposal ID.
    }
\end{itemize}
In addition to these user-declared relations, 
there are reserved relations for built-in 
smart contract constructs that do not need to be declared.
For example, relation \lstinline{*msgSender(p: address)}
is reserved for incoming message sender address.
Similarly, there are reserved relations for message values, 
current block number, contract constructors, etc.

\noindent \textbf{Relation annotations.}
Three kinds of relation annotations are supported:
\begin{itemize}
    \item 
\lstinline{init} indicates that the relation is initialized
by a constructor argument passed during deployment.
    \item 
\lstinline{violation} means that the relation represents a 
safety violation query.
    \item 
\lstinline{public} generates a public interface to 
read the contents of the corresponding relational table.
\end{itemize}

\begin{figure}
\[
\boxed{
\begin{array}{rl}
    Rule & \coloneqq H(\bar{x}) \dldeduce Body \\
    Body & \coloneqq Join ~|~ R(\bar{x}), y = Agg: R(\bar{y}) \\
    Join & \coloneqq R(\bar{x}) ~|~ Pred, Join \\
    Agg & \coloneqq sum~n~|~max~n~|~min~n~|~count  \\
    Pred & \coloneqq R(\bar{x}) ~|~ C(\bar{x}) ~|~ y = F(\bar{x})\\
    (Condition)~C & \coloneqq > ~|~ < ~|~ \geq ~|~ \leq ~|~ \neq ~|~ == \\ 
    (Function)~F & \coloneqq + ~|~ - ~|~ \times ~|~ / \\
\end{array}
}
\]
    \caption{Syntax of DeCon rules.
    $H(\bar{x})$ and $R(\bar{x})$ are relational literals,
    with $H$ and $R$ being the relation name, and $\bar{x}$ is an array
    of variables or constants.
    For the $max$, $min$, and $sum$ aggregators, $n$ is a variable
    in a numerical domain.}
    \label{fig:rule-syntax}
\end{figure}

\noindent \textbf{Rules.}
Figure~\ref{fig:rule-syntax} shows the syntax of DeCon rules.
A DeCon rule is of the form $head \dldeduce body$, 
which is interpreted from right to left: if the body is true,
then it inserts the head tuple into the corresponding relational table.

A rule body is a conjunction of literals,
and is evaluated to true if there exists a valuation of variables 
$\pi: V \mapsto D$ such that all literals are true.
$\pi$ maps a variable $v \in V$ to its concrete value in domain $D$.
Given a variable valuation $\pi$,
a relational literal is evaluated to true if and only if
there exists a matching row in the corresponding relational table.
Other kinds of literals, including conditions, functions,
and aggregations, are interpreted as constraints
on the variables.

In particular, DeCon supports three kinds of rules.
Their differences in syntax and semantics are described as follows:
\begin{itemize}
    \item 
    \textbf{Join rules} are rules that have a list of predicates in the rule body,
    and contain at least one relational literal.
    A predicate can be either a relational literal, 
    a condition, or a function.

    \item 
    \textbf{Transaction rules} are a special kind of join rules that 
    have one special literal in the body: transaction handlers.
    A transaction handler literal has \lstinline{recv_} prefix
    in its relation name, and is evaluated to true when
    the corresponding transaction request is received.
    The rest of the rule body specifies the approving condition for the transaction.

    \item \textbf{Aggregation rules} 
    are rules that contain a relational literal 
    $R(\bar{x})$ and an aggregator $y = Agg~n: R(\bar{y})$, 
    where $Agg$ can be either $max$, $min$, $count$, or $sum$.
    For each valid valuation of variables in $R(\bar{x})$,
    it computes the aggregate on the matching rows in $R(\bar{y})$.
    Take the following rule from the voting contract as an example.
    \begin{lstlisting}
votes(p,c) :- vote (_,p), c = count: vote (_,p).
    \end{lstlisting}
    For each unique value $p$ in the second column of table \lstinline{vote},
    the aggregator \lstinline{c = count: vote (_,p)},
    counts the number of rows in table \lstinline{vote} whose second column 
    equals $p$.
\end{itemize}

DeCon contracts are executable on the Ethereum blockchain~\cite{chen2022declarative}.
To execute, they are first compiled into Solidity~\cite{solidity},
which is further compiled to bytecode for the Ethereum blockchain.
\section{Program Transformation}
\label{sec:program-transformation}

\begin{figure*}[ht]
\[
\infer[(Join1)]{
    \Gamma,\tau \vdash H(\bar{y}) \dldeduce R(\bar{x}) \hookrightarrow \phi
}{
    \Gamma,\tau \vdash R(\bar{x}) \rightsquigarrow \phi
}
~~\infer[(Join2)]{
    \Gamma,\tau \vdash H(\bar{y}) \dldeduce Pred, Join \hookrightarrow \phi_1 \land \phi_2
}{
    \Gamma,\tau \vdash Pred \rightsquigarrow \phi_1
    & 
    \Gamma,\tau \vdash H(\bar{y}) \dldeduce Join \hookrightarrow \phi_2
}
\]
\[
\infer[(Sum+)]{
    \Gamma,\tau \vdash H(\bar{y}) \dldeduce R(\bar{x}), s=sum~n: R(\bar{z})
        \hookrightarrow s=s'
}{
    \tau = insert \_
    & s'=\Gamma(H)[\bar{k}].value + n
}
\quad
\infer[(Sum-)]{
    \Gamma,\tau \vdash H(\bar{y}) \dldeduce R(\bar{x}), s=sum~n: R(\bar{z})
        \hookrightarrow s=s'
}{
    \tau = delete \_
    & s'=\Gamma(H)[\bar{k}].value - n
}
\]
\[
\infer[(Count+)]{
    \Gamma,\tau \vdash H(\bar{y}) \dldeduce R(\bar{x}), c=count: R(\bar{z}) 
        \hookrightarrow \phi
    }{
      \tau = insert \_
     &   \phi \coloneqq c=\Gamma(H)[\bar{k}].value+1
    }
\quad
\infer[(Count-)]{
    \Gamma,\tau \vdash H(\bar{y}) \dldeduce R(\bar{x}), c=count: R(\bar{z}) 
        \hookrightarrow \phi
    }{
     \tau = delete \_
    &   \phi \coloneqq c=\Gamma(H)[\bar{k}].value-1
    }
\]
\[
\infer[(Max)]{
    \Gamma, \tau \vdash H(\bar{y}) \dldeduce R(\bar{x}), m=max~n: R(\bar{z}) 
        \hookrightarrow \phi
    }
    { 
     \tau = insert~\_
     & m' = \Gamma(H)[\bar{k}].value
    & \phi \coloneqq (n > m' \land m=n) \oplus (n \leq m') 
    }
\]
\[
\infer[(Min)]{
    \Gamma, \tau \vdash H(\bar{y}) \dldeduce R(\bar{x}), m=min~n: R(\bar{z})
        \hookrightarrow  \phi
    }
    {
     \tau = insert~\_
     & m' = \Gamma(H)[\bar{k}].value
    & 
    \phi \coloneqq (n < m' \land m=n) \oplus (n \geq m')
    }
\]
\caption{Inference rules for the EncodeRuleBody procedure.}
\label{fig:inference-rules}
\end{figure*}

\subsection{Declarative Smart Contracts as Transition Systems}

This section introduces the algorithm to translate a DeCon smart contract
into a state transition system $\langle S,I,E,Tr \rangle$ where
\begin{itemize}
    \item $S$ is the state space: the set of all possible valuations 
        of all relational tables in DeCon.
    \item $I \subseteq S$ is the set of initial states that 
        satisfy the initial constraints of the system.
        All relations are by default initialized to zero, or unconstrained
        if they are annotated to be initialized by constructor arguments.
    \item $E$ is the set of transaction types. Each element in $E$ corresponds to a
        type of transaction in DeCon 
        (analogous to a transaction function definition in Solidity).
    \item $Tr \subseteq S \times E \times S$ is the transition relation, 
        generated from DeCon rules.
        $Tr(s,e,s')$ means that state $s$ can transit to state $s'$
        via transaction $e$.
\end{itemize}

In the rest of this section, we introduce the algorithm to generate the transition
relation from a DeCon smart contract.

\subsection{Transition Relation} 
\label{sec:transition}

The transition relation $\var{Tr}$ is defined by
a formula $\var{tr}: S \times E \times S \mapsto Bool$.
Given $s,s' \in S, e \in E$, $s$ can transition to $s'$ in one step 
via transaction type $e$ if and only if $\var{tr}(s,e,s')$ is true.
Equation~\ref{eq:transition} defines $\var{tr}$ as a disjunction over 
the set of formulas encoding each transaction rule.
$R$ is the set of rules in the DeCon contract.
$TR$ is the set of transaction rules in $R$.
$\Gamma$ is a map from relation to its modeling variable,
e.g., the relation \lstinline{vote(proposal:uint,c:uint)[0]} 
is mapped to $\var{votes}: uint \mapsto uint$.
Recall from Section~\ref{sec:language} that transaction rules
are rules that listen to the incoming transaction and is only
triggered by the incoming transaction request. 
Therefore, $r.trigger$ is the literal with \lstinline{recv_} 
prefix in $r$'s body.

\begin{equation}
    \label{eq:transition}
    tr \triangleq \bigvee\limits_{r \in \var{TR}} [\text{EncodeRule}(r,R,\Gamma,r.trigger)
        \land e = r.TxName ]
\end{equation}


\begin{algorithm}
\caption{$\text{EncodeRule}(r,R,\Gamma,\tau)$.}
\label{alg:transform-rule}
\hspace*{\algorithmicindent} \textbf{Input}: 
     (1) A DeCon rule $r$, 
     (2) the set of all DeCon rules $R$,
     (3) a map from relation to its modeling variable $\Gamma$, 
     (4) a trigger $\tau$, the newly inserted literal that triggers $r$'s update.\\
\hspace*{\algorithmicindent} \textbf{Output}: 
    A formula over $S \times S$, encoding $r$'s body condition,
    and all state updates triggered by inserting $r$'s head literal.
\begin{algorithmic}[1]
    \State $ \var{Body} \gets \text{EncodeRuleBody}(\Gamma, \tau, r)$
    \State $ \var{Dependent} \gets \{ \text{EncodeRule}(dr,R,\Gamma, \tau')~|~
            (dr, \tau') \in \text{DependentRules}(r,R)\}$
    \State $(H, H') \gets \text{GetStateVariable}(\Gamma, r.head)$
    \State $\var{Update} \gets H' = \text{GetUpdate}(H, r, \tau)$
    \State $\var{TrueBranch} \gets \var{Body} \land \var{Update} \land (\bigwedge_{d \in \var{Dependent}} d ) $
    \State $\var{FalseBranch} \gets \neg \var{Body} \land (H'=H)$
    \State \Return $ \var{TrueBranch} \oplus \var{FalseBranch} $
\end{algorithmic}
\end{algorithm}

The procedure $\text{EncodeRule}$ is defined by Algorithm~\ref{alg:transform-rule}.
We explain it using the voting contract in Listing~\ref{lst:voting} as an example.

It takes four inputs: (1) a DeCon rule $r$,  
(2) the set of all DeCon rules $R$, 
(3) a map from relation to its modeling variable $\Gamma$, 
(4) and a trigger $\tau$ , the newly inserted literal that triggers $r$’s update.
In particular, a trigger $\tau$ takes the form
\lstinline{insert [literal]} or \lstinline{delete [literal]}.
It is used to inform the subroutine $\text{EncodeRuleBody}$ how 
a relation is updated, 
and that the rest of the rule body in $r$ needs to be encoded as 
a logical formula. 
For example, when a new \lstinline{vote} transaction is received,
the trigger $\tau$ is \lstinline{insert recv_vote(p)},
where \lstinline{p} is the transaction parameter.

In step 1, $r$'s body is encoded as a boolean formula, $\var{BodyConstraint}$, 
by calling a procedure $\var{EncodeRuleBody}$ (Section~\ref{sec:encode-rule-body}).
Take the rule for \lstinline{vote} transaction in line 15 of Listing~\ref{lst:voting} as an example.
Its body is encoded as:
\[      \neg \var{hasWinner}  
      \land~ \neg \var{hasVoted}[v]  
      \land~ \var{isVoter}[v]  \]


Step 2 first selects direct dependent rules of $r$ from the set of 
all DeCon rules $R$, by calling a subroutine $\var{DependentRules(r,R)}$. 
It returns a set of tuple $(dr,\tau')$, where $dr$ is a direct dependent rule of $r$,
and $\tau'$ is the corresponding trigger for $dr$.
A rule $dr$ is directly dependent on rule $r$ if and only if $r$'s head relation
appears in $dr$'s body. For example, rules in line 19 and 26 of Listing~\ref{lst:voting} are directly dependent
on the \lstinline{vote} transaction rule in line 15.
For the next trigger $\tau'$,
literal insertion results from a new relational tuple being derived 
from one of the rules.
For example, if the rule for transaction \lstinline{vote} 
is evaluated to true,
the next trigger $\tau'$ is \lstinline{insert vote(v,p)}.
Literal deletion happens when literals with primary keys are inserted: 
inserting such literals implicitly deletes the literals
with the same primary keys, if exist.
Next, for each direct dependent rule $dr$ of $r$ and trigger $\tau'$, 
it gets $dr$'s encoding by recursively calling itself on $dr$ and $\tau'$.

Step 3 generates state variables for the head relation,
where $H$ is for the current step, and $H'$ is for the next transition step.
Step 4 generates the head relation update constraint: 
$H'$ equals inserting or deleting $r$'s evaluation result from $H$.
$\var{GetUpdate(H,r,\tau)}$ is defined as follows:
\[
    \text{GetUpdate}(H,r,\tau)= 
\begin{cases}
    H.insert(r.head), & \\ ~~~~\text{if r is agg. rule} \lor \tau = \text{insert \_} \\
    H.delete(r.head), & \\ ~~~~\text{if r is join rule} \land \tau = \text{delete \_} \\
\end{cases}
\]
If $r$ is an aggregation rule, the update is directly encoded as insertion
since new aggregation results implicitly overwrite the old ones.
If $r$ is a join rule, and the trigger $\tau$ is a tuple deletion,
then $r$'s join result with the deleted tuple needs to be deleted as well.
Otherwise, $\tau$ is an insertion, and the update for relation $r.head$ is also an insertion.
Suppose we are in the recursion step for encoding 
the \lstinline{votes} rule in line 19, its update constraint 
is generated as:
$\var{votes}' = Store(\var{votes},p,\var{votes}[p]+1)$,
where the number of votes for proposal $p$ is incremented by one.

Step 5 generates the constraint where $r$'s body is true,
in conjunction with the update constraint and all dependent rules'
constraints. Step 6, on the other hand, generates constraints
where $r$'s body is false, no dependent rule is triggered,
and the head relation remains the same.
Step 7 returns the final formula as an exclusive-or of the true and
false branches,
which encodes $r$'s body and how its update affects
other relations in the contract.

\subsection{Encoding Rule Bodies}
\label{sec:encode-rule-body}
The procedure EncodeRuleBody is defined by two sets of inference rules:
\begin{itemize}
    \item 
$\Gamma, \tau \vdash r \hookrightarrow \phi$ 
states that a DeCon rule $r$ is encoded by a boolean formula $\phi$
under context $\Gamma$ and $\tau$.
\item
$ \Gamma,\tau \vdash \var{Pred} \rightsquigarrow \phi $
states that a predicate $\var{Pred}$ is encoded by
a formula $\phi$ under context $\Gamma$ and $\tau$.
\end{itemize}
The contexts ($\Gamma$ and $\tau$) of both judgments
are defined in the same way as the input of Algorithm~\ref{alg:transform-rule}.


Figure~\ref{fig:inference-rules} shows the inference rules
that define the first judgment 
$\Gamma, \tau \vdash r \hookrightarrow \phi$.
They are interpreted as follows.

A $Join$ rule is encoded as a conjunction of the predicates, 
each of which is encoded from a literal in the rule body.
The encoding of individual literals is introduced later in this section.

Unlike $Join$ rules,
aggregation rules ($Sum$ and $Count$) have separate inference rules for tuple
insertion ($+$) and deletion ($-$).
Because the relation between new and old aggregation results needs to be encoded.
In these reference rules,
$\bar{k}$ represents the primary keys of relation $H$, extracted from the
array $\bar{y}$, and $\Gamma(H)[\bar{k}].value$ reads the current aggregate result.
Note that, unlike the $Join$ rules, the literal $R(\bar{x})$ 
here does not join with the aggregation literal, 
because it is only introduced to obtain valid valuations for the rule variables
(every row in table $R$ is a valid valuation).
For each valid valuation, the aggregator computes the aggregate summary 
for the matching rows in table $R$ (Section~\ref{sec:language}).

For $Max$ and $Min$ aggregation rules, \proto only encodes the their update 
for tuple insertions, based on the assumption that they only
apply to transaction relations (tables that stores the transaction records), 
which are append only and has no primary keys.
In other words, they have no tuple deletion.

This assumption is made for two reasons.
First, updating $Max$ and $Min$ for tuple deletion is complicated, because
if the current maximum or minimum is deleted, the second largest
or smallest element needs to be fetched and become the new aggregation result. 
Such update requires storing the whole table and even maintaining
sorted table entries.
Second, Ethereum has strict limits on the computation 
and storage of each smart contract and its transactions.
Maintaining maximum and minimum for tables with delete operation
is very expensive to be executed on Ethereum.
We survey smart contracts in public repositories and find no contract
with such logic.
Therefore, \proto adds such assumption and greatly simplify the rule encoding.

\noindent{\textbf{Encoding individual literals.}}
Following are the inference rules for judgment:
$ \Gamma,\tau \vdash \var{Pred} \rightsquigarrow \phi $,
which encodes individual literals.
\[
\infer[(Lit1)]{
    \Gamma, \tau \vdash R(\bar{x}) \rightsquigarrow \tau = R(\bar{x})
}{
    \tau.rel = R
  }
\]
\[
\infer[(Lit2)]{
  \Gamma, \tau \vdash R(\bar{x}) \rightsquigarrow \Gamma(R)[\bar{k}] = \bar{v}
}{
    \tau.rel \neq R
}
\]
where $\bar{k}$ represents the primary keys in relational literal $R(\bar{x})$, 
extracted from $\bar{x}$,
and $\bar{v}$ represents the remaining fields in $\bar{x}$.
\[
\infer[(Condition)]{
    \Gamma, \tau \vdash C \rightsquigarrow C
    }{}
\]
\[
\infer[(Function)]{
    \Gamma, \tau \vdash y = F(\bar{x}) \rightsquigarrow y=F(\bar{x})
    }{}
\]
Conditions and functions are directly encoded as they are,
as shown in the above rules.

\noindent{\textbf{Rule derivation and recursion.}}
\proto assumes that on every new incoming transaction request, 
there is at most one new tuple derived by each rule,
and that there is no recursion in the rules.

Rule recursion means that a rule is dependent on itself.
A rule $r_a$ is dependent to another rule $r_b$
($r_a \rightarrow r_b$)
if $r_b$'s head relation appears in $r_a$'s body.
This dependency relation is transitive:
$r_a \rightarrow r_b \land r_b \rightarrow r_c \implies r_a \rightarrow r_c$.
Using this dependency annotation ($\rightarrow$), rule recursion 
means $r_a \rightarrow ... \rightarrow r_a$.

This assumption keeps the size of the transition constraint 
linear to the number of rules in the DeCon contract,
thus making the safety verification tractable.
We find this assumption holds for most smart contracts 
in the financial domain, and is true for all of the ten
benchmark contracts in our evaluation.

\noindent{\textbf{Multi-contract Interactions.}}
Multi-contract interaction is specified implicitly by DeCon rules that join 
relations from different contracts.
Such interactions are performed via message passing.
Unlike prior work checking for message handling errors, 
DCV assumes that message delivery and handling are always successful,
and instead focuses on the functional correctness.
Note that such interactions are limited to functions without mutual recursions.
Mutual recursions are not supported because
it breaks the atomicity assumption of a transaction.
To illustrate, suppose
contract A's transaction calls contract B's transaction \textit{Foo},
which in turn calls another transaction of contract A.
In this case, the execution of two transactions of contract A overlap,
breaking the atomicity of transactions.

\subsection{Safety Invariant Generation}

Each violation query rule $qr$ in a DeCon contract 
is first encoded as a formula $\phi$ such that
$\Gamma,\tau \vdash qr \hookrightarrow \phi$.
Note that the context $\Gamma$ is the same mapping used 
in the transition system encoding process.
The second context, trigger $\tau$, is a reserved literal $check()$,
which triggers the violation query rule after every transaction.

Next, the safety invariant is generated from $\phi$ as follows:
\[
    Prop \triangleq \neg (\exists x \in X.~ \phi(s,x))
\]
where $X$ is the state space for the set of non-state variables in $\phi$.
The property states that there exists no valuations of variables
in $X$ such that the violation query is non-empty.
In other words, the system is safe from such violation.

\section{Verification Method}
\label{sec:verification-method}

\subsection{Proof by Induction}

Given the state transition system translated from the DeCon smart contract,
the target property $prop(s)$, which is translated from the violation query, 
is proven by mathematical induction.
In particular, let $S$ be the set of states in the transition system,
and $E$ be the set of transaction types (\lstinline{vote} is the only transaction
type in the example in Listing~\ref{lst:voting}).
Given $s,s'\in S, e \in E$, let $\var{init}(s)$ indicate whether $s$ is in the 
initial state, and $\var{tr}(s,e,s')$ indicate whether $s$ can transition
to $s'$ via transaction type $e$. The mathematical induction is as follows:

\begin{equation}
\label{eq:induction}
\boxed{
\begin{array}{rl}
\text{ProofInd}(init, tr, prop) \triangleq & \text{Base}(init,prop) \\ 
                                                   & \land \text{Induction}(tr,prop) \\
\text{Base}(init,prop) \triangleq & \forall s \in S.~init(s) \\
                                  & \implies inv(s) \land prop(s) \\
\text{Induction}(tr,prop) \triangleq & \forall s,s' \in S, e \in E.~ \\ 
                                     & inv(s) \land prop(s) \land tr(s,e,s') \\ 
                                     & \implies inv(s') \land prop(s') \\
\end{array}
}
\end{equation}
where $inv(s) \land prop(s)$ is an inductive invariant inferred by \proto 
such that $prop(s)$ is proved to be an invariant of the transition system.

\begin{algorithm}
\caption{Procedure to find inductive invariants.}
\label{alg:find-indcutive-invariant}
\hspace*{\algorithmicindent} \textbf{Input}: 
    a transition system $ts$, 
    a map from relation to its modeling variable $\Gamma$,
    and a set of DeCon transaction rules $R$. \\
\hspace*{\algorithmicindent} \textbf{Output}: 
    an inductive invariant of $ts$.
\begin{algorithmic}[1]
    \Function{FindInductiveInvariant}{C,ts}
        \For{inv \textbf{in} C}:
            \If{refuteInv(inv, C, ts)}
                \State \Return $\text{FindInductiveInvariant}(C \setminus inv, ts)$
            \EndIf
        \EndFor
        \State \Return $ \bigwedge_{c_i \in C} c_i $
    \EndFunction
    \State $P \gets \bigcup_{r \in R} \text{ExtractPredicates}(r,\Gamma) $
    \State $C \gets \text{GenerateCandidateInvariants(P)} $ 
    \State \Return $\text{FindInductiveInvariant}(C,ts)$
\end{algorithmic}
\end{algorithm}

Algorithm~\ref{alg:find-indcutive-invariant} presents the procedure to 
infer inductive invariants.
It first extracts a set of predicates $P$ from the set of transaction rules $R$
(Section~\ref{sec:extract}).
Then it generates a set of candidate invariants using predicates in $P$,
following two heuristic patterns (Section~\ref{sec:patterns}).
Finally, it invokes a recursive subroutine \textsc{FindInductiveInvariant}
to find an inductive invariant.

The procedure \textsc{FindInductiveInvariants} is adopted from the Houdini algorithm~\cite{lahiri2009complexity}. 
It iteratively refutes candidate invariants in $C$, until there is no candidate that  
can be refuted, and returns the conjunction of all remaining invariants.
The subroutine \text{refuteInv} is defined in Equation~\ref{eq:refute},
which refutes a candidate invariant if it is not inductive:
\begin{equation}
\begin{array}{rl}
    \text{refuteInv}(inv,C,ts) \triangleq & \lor \neg (ts.init \implies inv) \\
        & \lor \neg [(\bigwedge_{c\in C}c) \land ts.tr \implies inv'] \\
\end{array}
\label{eq:refute}
\end{equation}
where $inv'$ is adopted by replacing all state variables in $inv$ 
with their corresponding variable in the next transition step.

A property of this algorithm is that, given a set of candidate invariants $C$,
it always returns the strongest inductive invariant that can be constructed 
in the form of conjunction of the candidates in $C$~\cite{lahiri2009complexity}.

\subsection{Predicate Extraction}
\label{sec:extract}

\begin{algorithm}
\caption{ExtractPredicate(r, $\Gamma$).}
\label{alg:extract-predicate}
\hspace*{\algorithmicindent} \textbf{Input}: 
    a transaction rule $r$, a map from relation to its modeling variable $\Gamma$. \\
\hspace*{\algorithmicindent} \textbf{Output}: 
    a set of predicates $P$.
\begin{algorithmic}[1]
    \State $\tau \gets r.trigger $
    \State $P_0 \gets \{ p ~|~ l \in r.body, \Gamma,\tau \vdash l \rightsquigarrow p \} $
    \State $P_1 \gets \{ p \land q ~|~ p \in P_0, q \in \text{MatchingPredicates}(p,r) \}$
    \State \Return $P_0 \cup P_1$
\end{algorithmic}
\end{algorithm}

Algorithm~\ref{alg:extract-predicate} presents the predicate extraction procedure. 
It first transforms each literal in the transaction rule into a predicate, 
and puts them into a set $P_0$.
Some predicates in $P_0$ do not contain enough information on their own, e.g.,
predicates that contain only free variables.
Because the logic of a rule is established on the relation among its literals
(e.g. two literals sharing the same variable $v$ means joining on the
corresponding columns).
On the contrary, predicates that contain constants, e.g. hasWinner == true,
convey the matching of a column to a certain concrete value, 
and can thus be used directly in candidate invariant construction.

Therefore, in the next step, each predicate $p$ in $P_0$ is augmented by one of its 
matching predicates in $\var{matchingPredicates}(p,r)$,
which is the set of predicates in rule $r$ that share at least one 
variable with predicate $p$. This set of augmented predicates is $P_1$.
Finally, the union of $P_0$ and $P_1$ is returned.

\subsection{Candidate Invariant Generation}
\label{sec:patterns}

Given the set of predicates in $P$, \proto generates candidate invariants 
following two heuristic template patterns. 
Each pattern is a logical formula that consists of one or more placeholders.
These placeholders are replaced by predicates extracted from transaction rules ($P$) 
during template instantiation. 
\proto unions all possible template instantiations as the set of candidate 
invariants.

The concrete forms of the two template patterns used in \proto are as follows:
\begin{gather*} 
  \{ \forall x \in X.~ \neg init(s) \implies \neg p(s,x) ~|~ p \in P \} \\
  \{ \forall x \in X.~ \neg init(s) \land q(s,x) \implies \neg p(s,x) ~|~ 
      p,q \in P \}
\end{gather*} 
where $X$ is the set of non-state variables in the body of the formula.
$\neg init(s)$ is used as the implication premise
so that the whole formula can be trivially implied by the 
transition system's initial constraints.

Having $\neg p$ as the implication conclusion is based on the
observation that, in order to prove safety invariants, a lemma is needed to
prevent the system from unsafe transitions.
Recall from Section~\ref{sec:extract} that predicate $p$ is one of the 
transaction conditions, thus having $\neg p$ in the invariant can prevent 
relevant transactions from committing.
In the second pattern, we add another predicate $q \in P_0$ in the implication
premise to make the pattern more robust to different contracts and properties.

\section{Evaluation}
\label{sec:eval}

\begin{table}[ht]
\fontsize{9}{11}\selectfont
\caption{Benchmark properties for group one.}
\label{tab:benchmarks}
\centering
\begin{tabular}{l|p{5.8cm}}
\hline
Benchmarks      & Properties                                                          \\
\hline
wallet          & No negative balance.                                                \\
crowFunding     & No missing fund.                                                    \\
ERC20           & Account balances add up to totalSupply.                             \\
ERC721          & All existing tokens have an owner.                                    \\
ERC777          & No default operator is approved for individual account.            \\
ERC1155         & Each token's account balances add up to that token's totalSupply.   \\
controllableToken & Account balances add up to totalSupply. \\
partitionToken & Account balances add up to totalSupply in each partition. \\
paymentSplitter & No overpayment.                                                     \\
vestingWallet   & No early release.                                                   \\
voting          & At most one winning proposal.                                       \\
auction         & Each participant can withdraw at most once.                        \\
\hline
\end{tabular}
\end{table}

\begin{table*}[ht]
\fontsize{9}{11}\selectfont
\caption{
    Verification efficiency measured in time (seconds). 
    TO stands for time-out after 1 hour. 
    Unknown ($?$) means the verifier cannot verify the contract property.
    Errors ($\times$) from solc are caused by a known software issue~\cite{solc-error}.
    Solc-verify also returns errors on some benchmarks because it
    fails to analyze part of the OpenZeppelin libraries used in those benchmarks.
    }
\label{tab:results}
\centering
\begin{tabular}{l|l|cc|c|cc|cc}
\hline
Group & Name      &\#Rules& LOC & \proto & \multicolumn{2}{|c|}{Solc} & \multicolumn{2}{c}{Solc-verify} \\
                &       &     &       & reference    & DeCon     & reference        & DeCon        \\
\hline
Open standards & wallet            & 12 & 67  & 1  & 1     & ? & 17      & ? \\
and examples & crowFunding       & 14 & 85  & 2  & 1     & $\times$    & ? & 38      \\
& ERC20             & 19 & 389 & 2  & 20    & $\times$   & $\times$   & 50      \\
& ERC721            & 13 & 520 & 2  & TO    & $\times$   & $\times$   & 50      \\
& ERC777            & 31 & 562 & 2  & TO    & ? & 74      & ? \\
& ERC1155           & 18 & 645 & 2  & 12    & TO      & 36      & 48      \\
& paymentSplitter   & 6  & 166 & 1  & TO    & 12      & 25      & ? \\
& vestingWallet     & 7  & 113 & 1  & TO    & ? & 55      & 14      \\
& voting            & 6  & 36  & 2  & $\times$ & TO      & ? & ? \\
& auction           & 13 & 146 & 58 & $\times$ & TO      & ? & ? \\
& controllableToken & 23 & 55  & 2  & 29    & 2       & $\times$   & 46      \\
& partitionToken    & 16 & 70  & 2  & 1     & 1       & 14      & 30      \\
\hline
Top ERC20 & bnb               & 24 & 172 & 2  & 3     & 1       & ? & 55      \\
Tokens & link              & 20 & 308 & 2  & 1     & 2       & 36      & 46      \\
& ltcSwapAsset      & 25 & 655 & 2  & TO    & $\times$   & $\times$   & 50      \\
& matic             & 25 & 510 & 2  & 2     & $\times$   & 74      & 60      \\
& shib              & 22 & 508 & 2  & 1     & $\times$   & 42      & 51      \\
& tether            & 27 & 474 & 2  & 42    & $\times$   & $\times$   & 49      \\
& theta             & 21 & 213 & 2  & 270   & 1       & 45      & 49      \\
& wbtc              & 28 & 731 & 2  & TO    & $\times$   & $\times$   & 67       \\
\hline
\end{tabular}
\end{table*}

\noindent \textbf{Implementation}.
We implement the smart contract transformation and inductive invariant
generation algorithms in Scala and use Z3~\cite{z3} 
to check the satisfiability of generated formulas. Quantified formulas are handled by Z3's 
default heuristics.

\noindent \textbf{Benchmarks}.
\new{
We collect 20 benchmark contacts in two groups.
The first group consists of 12 contracts from open 
libraries~\cite{openzepplin,solidity-examples}
and examples from prior research~\cite{verx-bench}.
Each selected contract either has contract-level safety specifications annotated, 
or has proper documentation from which we can come up with a 
contract-level safety specification.
Table~\ref{tab:benchmarks} shows the contract names in group one and their target properties.
The second group consists of the most popular contracts:
a list of top ERC20 contracts 
(based on circulating market cap) maintained by Etherscan~\cite{erc20-tokens}.
Based on this list,
we filter out the ones without source code or having unsupported features 
(e.g., cryptographic algorithms) and obtain another 8 benchmark contracts.
Since all of these contracts follow the ERC20 standards, they are verified against 
the same property for the ERC20 token contract.
}

\noindent \textbf{Baselines}.
We use solc~\cite{solidity} and solc-verify~\cite{hajdu2019solc} as the comparison
baselines.
Solc is a Solidity compiler with a built-in checker 
to verify assertions in source programs.
It has been actively maintained by the Ethereum community,
and version 0.8.13 is used for this experiment.
Solc-verify extends from solc 0.7.6 and performs automated formal verification 
using strategies of specification annotation and modular program verification.
We have also considered Verx~\cite{permenev2020verx} and Zeus~\cite{kalra2018zeus},
but neither is publicly available.

\noindent \textbf{Experiment setup}.
We modify certain functionalities and syntax of the benchmark contracts 
so that they are compatible with all comparison tools.
In particular, the delegate vote function of the voting contract contains
recursion, which is not yet supported by DeCon, and is thus dropped.
In addition, solc and solc-verify do not support inline assembly analysis.
Therefore, inline assembly in the Solidity contracts are replaced with native Solidity code. 
Minor syntax changes are also made to 
satisfy version requirements of the two baseline tools.

With these modifications, for each reference contract in Solidity, 
we implement its counterpart in DeCon.
Then we conduct verification tasks on three versions of benchmark contracts:
(1) DeCon contracts with \proto,
(2) reference Solidity contracts with solc and solc-verify,
and (3) Solidity contracts generated from DeCon with solc and solc-verify.
For each set of verification tasks, we measure the verification time
and set the time budget to be one hour.
All experiments are performed \new{on a 2.4GHz core (single-threaded) and 4GB memory}.

\noindent \textbf{Results}.
Table~\ref{tab:results} shows the evaluation results.
\new{\proto verifies all but one contract in two seconds,
with auction taking 58 seconds.
}
In particular, the properties for the voting and 
auction contract are not inductive, and thus require 
inductive invariant generation.
Auction takes more time because it contains more rules
and has a more complicated inductive invariant.

\new{
On the other hand, solc only successfully verifies 12 reference 
contracts, with one of them taking 270 seconds to finish. 
It times out on six contracts,
and reports SMT solver invocation error on another two.
}
This error has been an open issue according to the GitHub repository 
issue tracker~\cite{solc-error}, which 
is sensitive to the operating system and the underlying library versions of Z3.

Similarly, solc-verify verifies 10 reference contracts,
and reports unknown on four others. 
It also returns errors on six contracts
because it cannot analyze certain parts of the included OpenZepplin libraries,
although the libraries are written in compatible Solidity version.

For Solidity contracts generated from DeCon,
solc verifies six and solc-verify verifies 15.
The performance difference between the reference version
and the DeCon-generated version is potentially caused by the fact 
that DeCon generates stand-alone contracts that implement all 
functionalities without external libraries.
On the other hand, DeCon implements contract states (relations) 
as mappings from primary keys to tuples, which may incur 
extra analysis complexity compared to the reference version.



In summary, \proto is highly efficient in verifying 
contract-level safety invariants, and can handle a wider range of smart contracts compared to other tools. 
By taking advantage of the high-level abstractions of the DeCon
language, it achieves significant speedup over
the baseline tools. 
In several instances, baseline tools timeout after 
an hour or report an error, while DeCon is able to 
complete verification successfully.

\section{Related work}
\label{sec:related}

\noindent \textbf{Smart contract verification.}
Solc~\cite{marescotti2020accurate}, Solc-verify~\cite{hajdu2019solc}, 
Zeus~\cite{kalra2018zeus}, 
Verisol~\cite{wang2019formal}, 
and Verx~\cite{permenev2020verx}
perform safety verification for Solidity smart contracts.
Similar to \proto, they infer inductive invariants to perform sound
verification of safety properties.
They also generate counter-examples as a sequence of transactions
to disprove the safety properties.
SmartACE~\cite{wesley2022verifying} is a safety verification framework that
incorporates a wide variety of verification techniques, including fuzzing,
bounded model checking, symbolic execution, etc.
In addition to safety properties, SmartPulse~\cite{stephens2021smartpulse} 
supports liveness verification. It leverages 
the counterexample-guided abstraction refinement (CEGAR) paradigm to perform
efficient model checking, and can generate attacks given an environment 
model.

\proto differs from these work in that it uses a high-level executable
specification, DeCon, as the verification target.
Such high-level modeling improves verification efficiency,
but it also means that \proto can only apply to
smart contracts written in DeCon, 
while the other tools can work on most existing smart contracts
in Solidity or Move.

The Move Prover~\cite{dill2022fast} (MVP) is a formal verifier for
smart contracts written in the Move language~\cite{move}.
Similar to DCV, MVP also verifies safety properties. 
However, they target different languages and blockchain platforms. 
DCV is based on DeCon, which is declarative and more abstract,
while Move is imperative.
In addition, Move contracts work on the Diem blockchain, 
while DeCon currently supports Ethereum and Solidity.
Despite the differences, we believe DCV could also benefit Move. 
An interesting future direction would be to implement a Move compiler for DeCon, 
so that DeCon can serve as a declarative specification for smart contracts on the 
Diem blockchain, while Move as the implementation language can provide better 
support for other verification tasks.

\noindent \textbf{Formal semantics of smart contracts.}
KEVM~\cite{chen2018language} introduces formal semantics for smart contracts,
and can automatically verify that a Solidity program (its compiled EVM bytecode)
implements the formal semantics specified in KEVM.
This verification is also sound, but it focuses on the functional
correctness of each Solidity function, instead of the state 
invariants across multiple transactions.

Formal semantics of EVM bytecode have also been formalized
in F*~\cite{grishchenko2018semantic} and Isabelle/HOL~\cite{amani2018towards}.
Scilla~\cite{sergey2019safer} is a type-safe intermediate 
language for smart contracts that also provides formal semantics.
They offer precise models of the smart contract behaviors,
and support deductive verification via proof assistants.
However, working with a proof assistant requires non-trivial manual effort.
On the contrary, \proto provides fully automatic verification.

\noindent \textbf{Vulnerability detection.}
Securify~\cite{tsankov2018securify} 
encodes smart contract semantic information into relational facts,
and uses Datalog solver to search for property 
compliance and violation patterns in these facts.
Oyente~\cite{luu2016making} uses symbolic execution
to check generic security vulnerabilities,
including reentrancy attack, transaction order dependency, etc.
Maian~\cite{Nikoli2018finding} detects vulnerabilities by
analyzing transaction traces. 
Unlike the sound verification tools,
which require some amount of formal specification
from the users, these work require no formal 
specification and can be directly applied to 
any existing smart contracts without modification,
offering a quick and light-weight alternative to sound 
verification, although may suffer from false positives
or negatives.

\noindent \textbf{Fuzzing and testing.}
Fuzzing and testing techniques have also been widely applied 
to smart contract verification.
They complement deductive verification tools by
presenting concrete counter-examples.
ContractFuzzer~\cite{jiang2018contractfuzzer} instruments
EVM bytecodes to log run-time contract behaviors,
and uncovers security vulnerabilities from these 
run-time logs.
Smartisan~\cite{choi2021smartian} uses static 
analysis to predict effective transaction sequences,
and uses this information to guide fuzzing process.
SmartTest~\cite{so2021smartest} introduces a language 
model for vulnerable transaction sequences, and uses 
this model to guide the search path in the fuzzing phase.

\section{Conclusion}

We present \proto, an automatic safety verification tool
for declarative smart contracts written in the DeCon language. 
It leverages the high-level abstraction of DeCon to generate
succinct models of the smart contracts,
performs sound verification via mathematical induction,
and applies domain-specific adaptations of the Houdini algorithm
to infer inductive invariants. 
Evaluation shows that it is highly efficient, 
verifying all 20 benchmark smart contracts,
with significant speedup over the baseline tools.

Our experience with \proto has also inspired interesting directions
for future research. 
First, although \proto can verify a wide range of contracts
in the financial domain, we find certain interesting applications 
that require non-trivial extensions to the modeling language,
including contract inheritance, interaction between contracts,
and functions that lie outside relational logic.
Second, since \proto verifies on the contract logic-level,
we would also like to verify translation correctness for the 
DeCon-to-Solidity compiler, to 
ensure the end-to-end soundness of \proto's verification results.

\bibliographystyle{plain}
\bibliography{main}

\end{document}